\documentstyle[12pt]{article}
\input epsf
\catcode`\@=11
\def\marginnote#1{}
\def\ifmath#1{\relax\ifmmode #1\else $#1$\fi}

\def\bold#1{\setbox0=\hbox{$#1$}%
     \kern-.025em\copy0\kern-\wd0
     \kern.05em\copy0\kern-\wd0
     \kern-.025em\raise.0433em\box0 }

\def\GENITEM#1;#2{\par\vskip6pt \hangafter=0 \hangindent=#1
   \Textindent{$ #2$ }\ignorespaces}

\newcount\hour
\newcount\minute
\newtoks\amorpm
\hour=\time\divide\hour by60
\minute=\time{\multiply\hour by60 \global\advance\minute by-
\hour}
\edef\standardtime{{\ifnum\hour<12 \global\amorpm={am}%
    \else\global\amorpm={pm}\advance\hour by-12 \fi
    \ifnum\hour=0 \hour=12 \fi
    \number\hour:\ifnum\minute<100\fi\number\minute\the\amorpm}}
\edef\militarytime{\number\hour:\ifnum\minute<100\fi\number\minute}
\def\draftlabel#1{{\@bsphack\if@filesw {\let\thepage\relax
  \xdef\@gtempa{\write\@auxout{\string
    \newlabel{#1}{{\@currentlabel}{\thepage}}}}}\@gtempa
    \if@nobreak \ifvmode\nobreak\fi\fi\fi\@esphack}
     \gdef\@eqnlabel{#1}}
\def\@eqnlabel{}
\def\@vacuum{}
\def\draftmarginnote#1{\marginpar{\raggedright\scriptsize\tt#1}}
\def\draft{\oddsidemargin -.5truein
        \def\@oddfoot{\sl preliminary draft \hfil
        \rm\thepage\hfil\sl\today\quad\militarytime}
        \let\@evenfoot\@oddfoot \overfullrule 3pt
        \let\label=\draftlabel
        \let\marginnote=\draftmarginnote

\def\@eqnnum{(\theequation)\rlap{\kern\marginparsep\tt\@eqnlabel}%
\global\let\@eqnlabel\@vacuum}  }
\def\preprint{\twocolumn\sloppy\flushbottom\parindent 1em
        \leftmargini 2em\leftmarginv .5em\leftmarginvi .5em
        \oddsidemargin -.5in    \evensidemargin -.5in
        \columnsep 15mm \footheight 0pt
        \textwidth 250mmin      \topmargin  -.4in
        \headheight 12pt \topskip .4in
        \textheight 175mm
        \footskip 0pt

\def\@oddhead{\thepage\hfil\addtocounter{page}{1}\thepage}
        \let\@evenhead\@oddhead \def\@oddfoot{} \def\@evenfoot{}
}
\def\titlepage{\@restonecolfalse\if@twocolumn\@restonecoltrue\o
necolumn
     \else \newpage \fi \thispagestyle{empty}\c@page\z@
        \def\thefootnote{\fnsymbol{footnote}} }
\def\endtitlepage{\if@restonecol\twocolumn \else  \fi
        \def\thefootnote{\arabic{footnote}}
        \setcounter{footnote}{0}}  %\c@footnote\z@ }
\catcode`@=12
\relax
\def\be{\begin{equation}}
\def\ee{\end{equation}}
\def\bea{\begin{eqnarray}}
\def\eea{\end{eqnarray}}

\def\mst11{m_{\;\widetilde{t}_{1}}}

\def\mst22{m_{\;\widetilde{t}_{2}}}
\def\mst12{m_{\;\widetilde{t}_{1,2}}}

\def\msb11{m_{\;\widetilde{b}_{1}}}
\def\msb22{m_{\;\widetilde{b}_{2}}}
\def\msb12{m_{\;\widetilde{b}_{1,2}}}

\def\mwidetilde2{\widetilde{m}^{2}}

\relax
\begin{document}
\topmargin-1.cm
%\draft
%\preprint
%
\begin{titlepage}
\vspace*{-64pt}
\begin{flushright}
{\footnotesize
FERMILAB-Pub-97/433-A\\
CERN-TH/97-376\\
OUTP-97-77-P\\ 
PURD-TH-98-01\\
hep-ph/9801306 \\}
\end{flushright}

\vskip .7cm

\begin{center}
{\Large \bf  GUT Baryogenesis after Preheating: Numerical\\
 Study of the Production and Decay of $X$-bosons\\ }
\vskip .7cm

{\bf Edward W. Kolb$^{a,}$\footnote{E-mail:
     {\tt  rocky@rigoletto.fnal.gov}}},
{\bf Antonio Riotto$^{b,}$\footnote{E-mail:
     {\tt riotto@nxth04.cern.ch}}$^,$\footnote{
     On leave  from Department of Theoretical Physics,
     University of Oxford, U.K. }}
and {\bf Igor I. Tkachev$^{c,}$\footnote{E-mail:
{\tt tkachev@physics.purdue.edu}}}

\vskip.15in
{\it 
$^a$NASA/Fermilab Astrophysics Center \\ Fermilab
National Accelerator Laboratory, Batavia, Illinois~~60510-0500 \\ and\\
Department of Astronomy and Astrophysics\\
Enrico Fermi Institute,
University of Chicago, Chicago, Illinois~~60637-1433\\
\vspace{12pt}
$^b$Theory Division, CERN, CH-1211 Geneva 23, Switzerland\\
\vspace{12pt}
$^c$Department of Physics, Purdue University, West Lafayette, Indiana 47907\\ 
and\\
Institute for Nuclear Research of the Academy of Sciences of Russia\\
Moscow 117312, Russia
}

\end{center}

\vskip .5cm

\baselineskip=24pt

\begin{quote}
We perform a fully non-linear calculation of the production of
supermassive Grand Unified Theory (GUT) $X$ bosons during preheating,
taking into account the fact that they are unstable with a decay width
$\Gamma_X$. We show that parametric resonance does not develop if
$\Gamma_X$ is larger than about $10^{-2} m_X$. We compute the
nonthermal number density of superheavy bosons produced in the
preheating phase and demonstrate that the observed baryon asymmetry
may be explained by GUT baryogenesis after preheating if $\Gamma_X$ is
smaller than about $10^{-3} m_X$. \\ \\ PACS: 98.80.Cq
\end{quote}
\end{titlepage}

\setcounter{footnote}{0}
\setcounter{page}{0}
\newpage
%
% BODY
\baselineskip=20pt

\section{Introduction}
The horizon and flatness problems of the standard big-bang cosmology
are solved elegantly if during the evolution of the early universe the
energy density happened to be dominated by some form of vacuum energy
which resulted in a quasi-exponential growth of the scale factor
\cite{guth}.  An inflationary stage is also required to dilute any
undesirable remnants such as topological defects surviving from some
phase transition at a pre-inflation epoch.

The vacuum energy driving inflation is generally assumed to be
associated with some scalar field $\phi$, known as the {\it inflaton,}
which is initially displaced from the minimum of its potential. As a
by-product of solving the horizon and flatness problem, quantum
fluctuations of the inflaton field may produce the seeds necessary for
the generation of structure formation and for fluctuations in the
cosmic background radiation.

Inflation ended when the potential energy associated with the inflaton
field became smaller than the kinetic energy of the field.  By that
time, any pre-inflation entropy in the universe had been inflated
away, and the energy of the universe was entirely in the form of
coherent oscillations of the inflaton condensate around the minimum of
its potential.  The universe may be said to be frozen after the end of
inflation. We know that somehow the low-entropy cold universe
dominated by the energy of coherent motion of the $\phi$ field must be
transformed into a high-entropy hot universe dominated by
radiation. The process by which the energy of the inflaton field is
transferred from the inflaton field to radiation has been dubbed
reheating.\footnote{Reheating may well be a misnomer since there is no
guarantee that the universe was hot before inflation.  Since we are
confident that the universe was frozen at the end of inflation,
perhaps ``defrosting'' is a better description of the process of
converting inflaton coherent energy into entropy.}

In the old theory of reheating \cite{old}, it was assumed that the
inflaton field oscillated around the minimum of its potential in a
coherent way until the age of the universe grew to the order of the
inflaton decay lifetime $\tau_\phi$, $t\sim\tau_\phi =
\Gamma_\phi^{-1}$. At this stage, the inflaton decayed and the
universe filled with the inflaton decay products, which soon
thermalized.  In the process the universe was ``reheated'' to the
temperature of $T_{RH}\simeq 10^{-1} \sqrt{\Gamma_\phi M_{\rm Pl}}$,
where $M_{\rm Pl}\sim 10^{19}$GeV is the Planck mass. In a simple
chaotic inflation model the inflaton potential is given by $V(\phi) =
m^2\phi^2/2$, with $m\sim 10^{13}$GeV in order to reproduce the
observed temperature anisotropies in the microwave background.
Writing $\Gamma_\phi=\alpha_\phi m$, one finds $T_{RH}\simeq
10^{15}\sqrt{\alpha_\phi}$ GeV \cite{ktbook}.

The density and temperature fluctuations observed in the present
universe, $\delta\rho/\rho\sim 10^{-5}$, require the inflaton
potential to be extremely flat. This means that the couplings of the
inflaton field to the other degrees of freedom (including
$\alpha_\phi$) cannot be too large, since large couplings would induce
large loop corrections to the inflaton potential, spoiling its
flatness. As a result, $T_{RH}$ is expected to be smaller than
$10^{14}$GeV by several orders of magnitude. The problem of large loop
corrections to the inflaton potential may be solved in the framework
of supersymmetry \cite{susy}, where the nonrenormalization theorem
\cite{th} guarantees that the superpotential is not renormalized at
any order of perturbation theory. On the other hand, in
supergravity-inspired scenarios gravitinos have a mass of order a TeV
and a decay lifetime on the order of $10^5$s. If gravitinos would be
overproduced in reheating and decay after the epoch of
nucleosynthesis, they would modify the successful predictions of
big-bang nucleosynthesis.  This can be avoided if the reheat
temperature is smaller than about $10^{11}$GeV (or even less,
depending on the gravitino mass).

In addition to entropy, the baryon asymmetry must be created
after inflation.  One method to generate the baryon asymmetry is by
the decay of baryon-number ($B$) violating superheavy bosons (referred
to generically as ``$X$'' bosons, whether gauge or Higgs bosons) of
Grand Unified Theories (GUT's) \cite{review}. In the old theory of
reheating there is a serious obstacle to post-inflation GUT
baryogenesis related to the relatively large $X$-boson mass and the
relatively small reheat temperature.

The unification scale is generally assumed to be around $10^{16}$GeV,
and $B$-violating gauge bosons should have masses comparable to this
scale. Baryon-number violating Higgs bosons may have a mass one or two
orders of magnitude less.  For example, in $SU(5)$ there are $B$
violating ``Higgs'' bosons in the five-dimensional representation that
may have a mass as small as $10^{14}$GeV. In fact, these Higgs bosons
are more likely than gauge bosons to produce a baryon asymmetry since
it is easier to arrange the requisite CP violation in the Higgs decay
\cite{KW,FOT,CP}.  But even the light $B$-violating Higgs bosons are
expected to have masses larger than the inflaton mass, and it would be
kinematically impossible to create them directly in $\phi$ decay.

One might think that the $X$ bosons could be created by thermal
scattering during the stage of thermalization of the decay products of
the inflaton field. However, if $T_{RH}$ is as small as necessary to
avoid overproduction of gravitinos, production of superheavy bosons by
thermal scattering would be heavily suppressed.\footnote{There exists
another problem for GUT baryogenesis scenarios: $B$ violation through
sphaleron transitions are expected to be fast at high temperatures,
and would erase any preexisting baryon asymmetry produced at the GUT
scale \cite{sh} unless there is a non vanishing value of $B-L$.  But a
natural way to overcome this problem is to adopt a GUT like $SO(10)$,
where an asymmetry in $B-L$ may be generated.}

But the outlook for GUT baryogenesis has brightened recently with the
realization that reheating may differ significantly from the simple
picture described above \cite{preheating1,KT1,KT2,KT3,KT4}.  In the
first stage of reheating, called ``preheating'' \cite{preheating1},
nonlinear quantum effects may lead to an extremely effective
dissipational dynamics and explosive particle production even when
single particle decay is kinematically forbidden. Particles can be
produced in the regime of a broad parametric resonance, and it is
possible that a significant fraction of the energy stored in the form
of coherent inflaton oscillations at the end of inflation is released
after only a dozen or so oscillation periods of the inflaton. What is
most relevant for the present discussion is that preheating may play
an extremely important role for GUT generation of the baryon
asymmetry. Indeed, it was shown in \cite{klr} that the baryon
asymmetry can be produced efficiently just after the preheating era,
thus solving many of the problems that GUT baryogenesis had to face in
the old picture of reheating.

A crucial observation for baryogenesis is that even particles with
mass larger than that of the inflaton may be produced during
preheating. To see how this might work, let us assume that the
interaction term between the superheavy bosons and the inflaton field
is of the type $g^2\phi^2|X|^2$.  During preheating, quantum
fluctuations of the $X$ field with momentum $\vec{k}$ approximately
obey the Mathieu equation: $ X_k'' + [A(k) - 2q\cos2z]X_k =0$, where
$q = g^2 \phi^2 / 4 m^2$, $A(k) = (k^2 + m_X^2) / m^2 + 2q$, and
primes denotes differentiation with respect to $z=m t$.  Particle
production occurs above the line $A = 2 q$.  The width of the
instability strip scales as $q^{1/2}$ for large $q$, independent of
the $X$ mass.  The condition for broad resonance, $A-2q {\
\lower-1.2pt\vbox{\hbox{\rlap{$<$}\lower5pt\vbox{\hbox{$\sim$}}}}\ }
q^{1/2}$ \cite{preheating1}, becomes $(k^2 + m^2_X)/m^2 {\
\lower-1.2pt\vbox{\hbox{\rlap{$<$}\lower5pt\vbox{\hbox{$\sim$}}}}\ } g
\bar\phi / m$, which yields for the typical energy of $X$ bosons
produced in preheating $E_X^2 = {k^2 + m^2_X } {\
\lower-1.2pt\vbox{\hbox{\rlap{$<$}\lower5pt\vbox{\hbox{$\sim$}}}}\ } g
\bar\phi m $ \cite{KT2}, where $\bar\phi$ is the amplitude of the
oscillating inflaton field.  By the time the resonance develops to the
full strength, $\bar\phi^2 \sim 10^{-5} M_{\rm Pl}^2$.  The resulting
estimate for the typical energy of particles at the end of the broad
resonance regime for $m \sim 10^{-6} M_{\rm Pl}$ is $E_X \sim 10^{-1}
g^{1/2}\sqrt { m M_{\rm Pl}} \sim g^{1/2} 10^{15}$ GeV.  Supermassive
$X$ bosons can be produced by the broad parametric resonance for $E_X
> m_X$, which leads to the estimate that $X$ production will be
possible if $m_X < g^{1/2} 10^{15}$ GeV.

For $g^2 \sim 1$ one would have copious production of $X$ particles
(in  this regime the problem
is non-linear from the beginning and therefore $g^2=1$ has to
 be understood as a rough
estimate of the limiting case) as heavy as $10^{15}$GeV, i.e., 100
times greater than the inflaton mass.\footnote{In the case in which
the cross-coupling between the inflaton and the $X$ field is negative,
superheavy particles may be produced even more efficiently \cite{pr}.}
The only problem here is that for large coupling $g$, radiative
corrections to the effective potential of the inflaton field may
modify its shape at $\phi \sim M_{\rm Pl}$.  However, this problem
does not appear if the flatness of the inflaton potential is protected
by supersymmetry.

This is a significant departure from the old constraints of reheating.
Production of $X$ bosons in the old reheating picture was
kinematically forbidden if $m< m_X$, while in the new scenario it is
possible because of coherent effects. It is also important to note
that the particles are produced out-of-equilibrium, thus satisfying
one of the basic requirements to produce the baryon asymmetry
\cite{sak}.

Scattering of X fluctuations off the zero mode of the inflaton field
limits the maximum magnitude of X fluctuations to be $\langle
X^2\rangle_{\rm max} \approx m^2/g^2$ \cite{KT3}.  For example,
$\langle X^2\rangle_{\rm max} \sim 10^{-10} M_{\rm Pl}^2$ in the case
$m_X = 10\:m$. This restricts the corresponding number density of
created $X$-particles.

A potentially important dynamical effect is that the parametric
resonance is efficient only if the self-interaction couplings of the
superheavy particles are not too large.  Indeed, a self-interaction
term of the type $\lambda |X|^4$ provides a non-thermal mass to the
$X$ boson of the order of $(\lambda \langle X^2\rangle)^{1/2}$, but
this contribution is smaller than the bare mass $m_X$, if $\lambda{\
\lower-1.2pt\vbox{\hbox {\rlap{$<$}\lower5pt\vbox{\hbox{$\sim$}}}}\ }
g^2 m_X^2/m^2$.  Self-interactions may also terminate the resonance
effect because scattering induced by the coupling $\lambda$ may remove
particles from the resonance shells and redistribute their momenta
\cite{klr,pr1}. But this only happens if, again, $\lambda \gg g^2$
\cite{KT2}.

The parametric resonance is also rendered less efficient when the $X$
particles have a large decay width $\Gamma_X$.  Roughly speaking, one
expects that the explosive production of particles takes place only if
the typical time, $\tau_e$, during which the number of $X$ bosons
grows by a factor of $e$, is smaller than the decay lifetime
$\tau_X=\Gamma_X^{-1}$.  During the broad resonance regime, typically
$\tau_e {\ \lower-1.2pt\vbox{\hbox
{\rlap{$>$}\lower5pt\vbox{\hbox{$\sim$}}}}\ }10\:m^{-1}$. If we
parameterize the decay width by $\Gamma_X=\alpha\:m_X$, this requires
$\alpha {\ \lower-1.2pt\vbox{\hbox
{\rlap{$<$}\lower5pt\vbox{\hbox{$\sim$}}}}\ } 0.1m/m_X$ \footnote{See also ref. \cite{kasuya}.}.  Notice that
smaller values of $\Gamma_X$ are favored not only because particle
production is made easier, but also because the superheavy particles
may remain out-of-equilibrium for longer times, thus enhancing the
final baryon asymmetry.

The exact knowledge of the maximum allowed value of the decay width of
the superheavy degrees of freedom is therefore of extreme importance
for the computation of the final baryon asymmetry produced by the GUT
particles after preheating.

The goal of this paper is twofold. First, we wish to provide the first
fully non-linear calculation of the inflaton decay into superheavy $X$
bosons taking into account their decay width $\Gamma_X$. Our basic
finding is that the parametric resonance does not develop if the decay
rate $\Gamma_X$ is larger than about $10^{-1}\:m$, thus confirming the
rough estimate made above.  $X$ production through the resonance is
very efficient for smaller values of $\Gamma_X$.  Our second goal is
to compute numerically the number density $n_X$ of supermassive $X$
bosons produced at the resonance stage. This parameter is fundamental
for the computation of the final baryon asymmetry.  We will also show
that as long as the bound $\Gamma_X {\ \lower-1.2pt\vbox{\hbox
{\rlap{$<$}\lower5pt\vbox{\hbox{$\sim$}}}}\ } 10^{-2}m$ is satisfied,
the observed baryon asymmetry $B\simeq 4\times 10^{-11}$ may be
explained by the phenomenon of GUT baryogenesis after preheating, with
no further restriction of the parameters. We will also comment on the
phenomenological implications of our findings.

\section{$X$ production and decay}
Using the methods developed in Refs.\ \cite{KT1,KT2,KT3}, we have
studied numerically the production of massive, unstable $X$ particles
in the process of the inflation decay.

We consider a model in which the oscillating inflaton field $\phi$
interacts with a scalar field $X$ whose decays violate baryon number
$B$. A simple possibility for the $X$-particle is the Higgs field in
the five-dimensional representation of $SU(5)$, although as noted
above, because of the desirability of $B-L$ violation $SO(10)$ is a
more promising theory. We assume standard kinetic terms, minimal
coupling with gravity, and a very simple potential for the fields of
the form
\begin{equation}
V(\phi,X)={1\over 2} m^2\phi^2  + {1\over 2} m_X^2 X^2 +
{1\over 2} g^2 \phi^2 X^2  \,   .
\label{pot}
\end{equation}

Let us first introduce dimensionless variables in the conformal
reference frame.  The rescaled conformal time $\tau$ is related to
cosmological time $t$ by $m dt= a(\tau)d\tau$.  The rescaled conformal
fields $\chi$ and $\varphi$ are related to the original fields by
$X(\tau)=\chi(\tau)\phi_0(0)/a(\tau)$ and $\phi(\tau) =\varphi(\tau)
\phi_0(0)/a(\tau)$.  In this model $\phi_0(0)\approx 0.28 M_{\rm Pl}$
is the value of the inflaton field at the end of inflation.  We assume
that immediately after inflation the universe is matter dominated and
the scale factor evolves as $a(\tau)=
(\sqrt{\pi}\phi_0(0)\tau/\sqrt{3}M_{\rm Pl} + 1)^2$ \cite{KT2}.

An important dimensionless parameter in the problem will be the
resonance parameter $q=g^2\phi^2_0(0)/4m^2$.  For $m =1.3\times
10^{-6}M_{\rm Pl}$ and $\phi_0(0) = 0.28 M_{\rm Pl}$, $q \simeq
10^{10}g^2$.

In the conformal variables, the equations of motion become
\begin{eqnarray}
{\ddot \varphi} - \nabla^2 \varphi  + \left(a^2-\frac{\ddot{a}}{a}
\right)\varphi + 4q\chi^2\varphi&=& 0 \; ,
\nonumber \\
{\ddot \chi} - \nabla^2 \chi + \Gamma a \dot{\chi} +
\left(m_\chi^2 a^2- \Gamma \dot{a}-\frac{\ddot{a}}{a} \right) \chi
+ 4q\varphi^2\chi&=& 0 \; .
\label{eqm}
\end{eqnarray}
Here we are taking into account the decay of the $X$ field by simply
introducing the term $\Gamma_X\, \dot{X}$ in the equation of motion
for the field $X$. The dimensionless parameter $\Gamma$ which enters
the Eq.\ (\ref{eqm}) is $\Gamma \equiv \Gamma_X/m$. Similarly, $m_\chi
= m_X/m$.

We have solved these equations of motion directly in coordinate
space on a $128^3$ spatial lattice with periodic boundary
conditions. The initial conditions for fluctuations correspond to the
conformal vacuum at the time when the oscillations of $\varphi_{0}$
commence \cite{KT1,KT2}.  The initial conditions for the coherently
oscillating inflaton zero-momentum mode are $\varphi_0 (0) =1$,
$\dot{\varphi}_0(0) =0$.

A fundamental parameter in GUT baryogenesis is $n_X$, the number
density of the supermassive leptoquarks whose decays produce the
baryon asymmetry.  It will depend upon the value of $\Gamma$ and $q$.

Since the supermassive bosons are more massive than the inflaton, one expects 
small kinetic energy in the excitations of the $X$ field.  From the potential 
of Eq.\ (\ref{pot}), the square of the
effective mass of the $X$ field is 
$(m^{\rm  EFF}_X)^2 = (m_X^2+g^2\langle \phi^2\rangle)$ and the
energy density in the
$X$ field will be 
$\rho_X=(m_X^2+g^2\langle \phi^2\rangle)\langle X^2\rangle$.
Writing $\langle \phi^2\rangle$ as $\phi_0^2 + \langle \delta\phi^2\rangle$,
we can define an analog of the $X$-particle number density as
\begin{equation}
n_X=\rho_X/m^{\rm  EFF}_X = \left[4q(\phi_0^2 +
\langle\delta\phi^2\rangle)/\phi_0^2(0) +
m_{\chi}^2\right]^{1/2} m \langle X^2 \rangle \, .
\label{n}
\end{equation}

Eq.\ (\ref{n}) enables us to calculate the number density of the
created $X$-particles if the variances of the fields, $\langle
X^2\rangle$, $\langle\delta\phi^2\rangle \equiv \langle
[\phi(x)-\langle\phi\rangle]^2 \rangle$, and the inflaton zero mode
$\phi_0 (\tau) \equiv \langle\phi(\tau)\rangle$, are known.  Here
$\langle \cdots \rangle$ has to be understood as the average over
statistical realizations.  Since the system is homogeneous on average,
this is equivalent to the volume average.  We shall present the
dependence upon the time of the variances for several choices of the
model parameters as well as the maximum value of the variances which
may be achieved during the evolution as a function of the same
parameters.

%%%%%%%%%%%%%%%%%%%%%%%%%%%%%%%%%%
%%%%%%%%%%%%%%%%%%%%%%%%%%%%%%%%%%
%%%%%%%%%%%%%%%%%%%%%%%%%%%%%%%%%%
 \begin{figure}
\centering
\leavevmode\epsfysize=3.2in \epsfbox{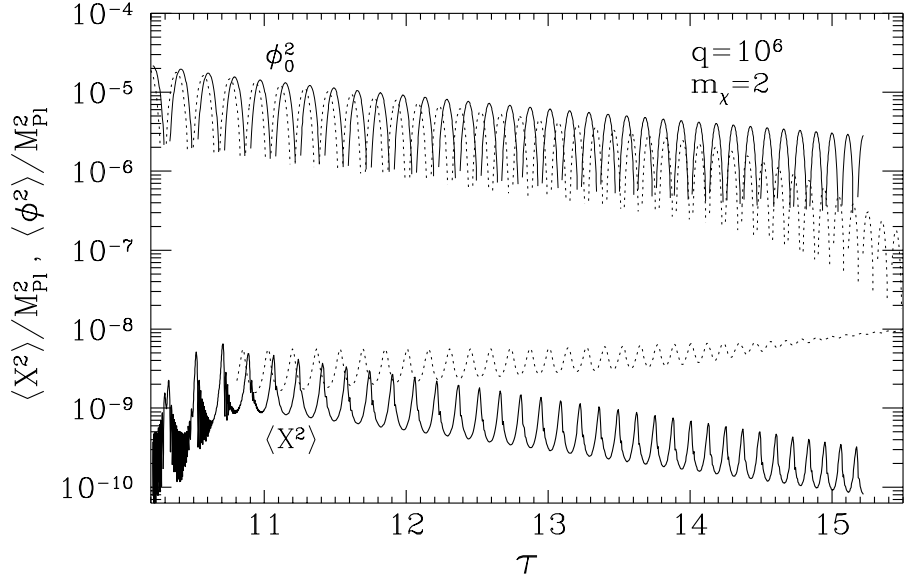}
\caption{The variance of $X$ with model parameters $q=10^6$, $m_\chi=2$, and
$\Gamma=6\times10^{-2}$ is shown by the lower solid curve as a
function of time.  The upper solid curve corresponds to the inflaton
zero mode. The dotted curves represent the same quantities for $\Gamma
=0$.  }
\label{fig:Fig1}
\end{figure}
%%%%%%%%%%%%%%%%%%%%%%%%%%%%%%%%%%
%%%%%%%%%%%%%%%%%%%%%%%%%%%%%%%%%%
%%%%%%%%%%%%%%%%%%%%%%%%%%%%%%%%%%
%%%%%%%%%%%%%%%%%%%%%%%%%%%%%%%%%%
%%%%%%%%%%%%%%%%%%%%%%%%%%%%%%%%%%
%%%%%%%%%%%%%%%%%%%%%%%%%%%%%%%%%%
\begin{figure}
\centering
\leavevmode\epsfysize=3.2in \epsfbox{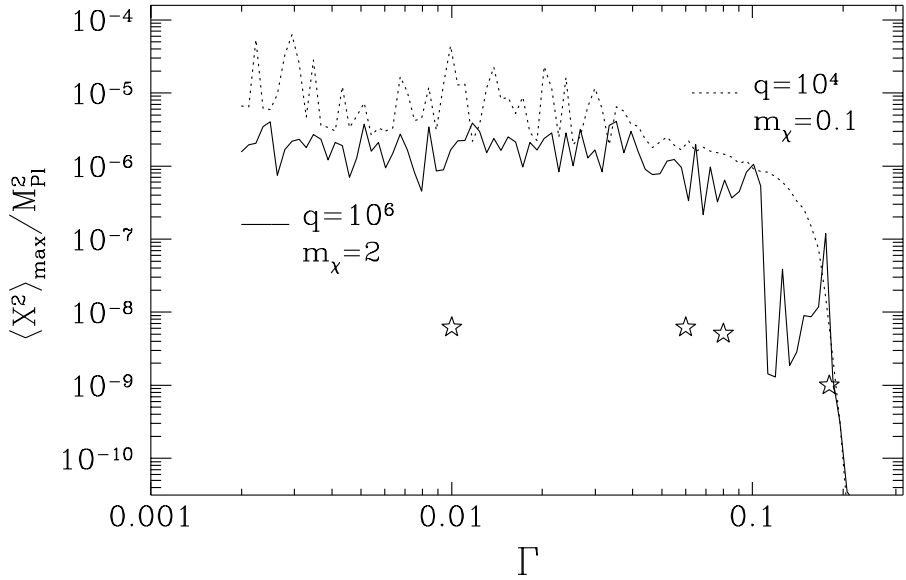}
%\leavevmode\epsfysize=7cm \epsfbox{scan.ps}
\caption{The maximum value of the variance of the $X$-field, $\langle
X_{\rm max}^2 \rangle$, is shown as function of $\Gamma$.  Stars mark
$\langle X_{\rm max}^2\rangle$ obtained in the full non-linear
problem. $\langle X_{\rm max}^2\rangle$ in the Hartree approximation
is shown by the dotted curve for $q=10^4$, $m_\chi=0.1$, and by the
solid curve for $q=10^6$, $m_\chi=2$.  }
\label{fig:Fig2}
\end{figure}
%%%%%%%%%%%%%%%%%%%%%%%%%%%%%%%%%%
%%%%%%%%%%%%%%%%%%%%%%%%%%%%%%%%%%
%%%%%%%%%%%%%%%%%%%%%%%%%%%%%%%%%%

The time evolution of the variance, $\langle X^{2} \rangle$, and of
the inflaton zero mode, $\langle \phi \rangle$, is shown in Fig.\
\ref{fig:Fig1} by the solid curves for the case $q=10^6$, $m_\chi=2$,
and $\Gamma=6\times10^{-2}$.  We see that the particle creation
reaches a maximum at $\tau \approx 10.8$ when $\langle X^{2} \rangle
\approx 10^{-9}$ in the ``valleys'' between the peaks.\footnote{Note
that we can use Eq.\ (\ref{n}) only for the ``valley'' values of the
variance, where the adiabatic approximation is valid, while
non-adiabatic amplification occurs in the region of the peaks of
$\langle X^{2}(\tau) \rangle$ \cite{KT3}.}  At later times, $\tau >
10.8$, particle creation by the oscillating inflaton field can no
longer compete with $X$-decays due to the non-zero value of
$\Gamma$. For comparison, we show in the same figure the case $\Gamma
=0$ represented by the dotted curves \cite{KT3}. In the $\Gamma =0$
case, particle creation is able to compete with the expansion of the
universe so that $\langle X^{2} \rangle$ remains roughly
constant. Another novel feature of the case with non-zero $\Gamma$ is
the low level of inflaton fluctuations, $\langle
\delta\phi^2\rangle(\tau ) \ll \phi_0^2(\tau )$.

Using Eq.\ (\ref{n}), we find for the maximum number density of created
$X$-particles $n_X=\left[4q\phi_0^2(10.8)/\phi_0^2(0) +
m_{\chi}^2\right]^{1/2} m \langle X^2 \rangle \approx \left[10^3 +
m_{\chi}^2\right]^{1/2} m \langle X^2 \rangle \approx 30 m \langle
X^2\rangle $.

It is easy to understand that if we increase the value of $\Gamma$, the
parametric resonance will not be able to compete with the decay of $X$
at earlier times.  Moreover, for sufficiently large values of
$\Gamma$, the resonance will be shut off in the linear
regime. One goal of this paper is to find the boundary of the model 
parameter space
that will result in sufficient $X$-particle creation for successful 
baryogenesis.

In exploration of parameter space we turn to the Hartree approximation
(for details see Ref.\ \cite {KT2}), which requires much less
computing resources. The maximum value of the variance of $X$ reached
during the time evolution of the fields in the Hartree approximation
is shown in Fig.\ \ref{fig:Fig2} as a function of the parameters of
the model. Here the stars also show the maximum of $\langle
X^{2}(\tau) \rangle$ in the full non-linear problem for a few values
of $\Gamma$.  At small $\Gamma$ the Hartree approximation
overestimates $\langle X^{2} \rangle$ significantly \cite{KT2,
KT3}. Nonetheless, at large values of $\Gamma$ it is a quite reliable
approach.  We see that $\langle X^{2} \rangle$ drops sharply when
$\Gamma > 0.2$, and we have checked that this critical value of
$\Gamma$ does not depend significantly upon $m_X$ or $q$.

The most relevant case with $q=10^8$, where $X$-bosons as massive as
ten times the inflaton mass can be created, is shown in Fig.\ 3 in the
Hartree approximation. Note, that two lower curves which correspond to
$\Gamma$ equal to 0.08 and 0.12 never reach the limiting value
$\langle X^2\rangle_{\rm max} \sim 10^{-10} M_{\rm Pl}^2$, which is
imposed by rescattering \cite{KT3}, and the Hartree approximation
ought to be reliable in this cases.

The final baryon asymmetry depends linearly upon the ratio $\delta$
between the energy stored in the $X$ particles at the end of the
preheating stage and the energy stored in the inflaton field at the
beginning of the preheating era \cite{klr}. From our results, we can
estimate that this ratio as
\begin{equation}
\delta\simeq 3\times 10^6\, \sqrt{\frac{q}{10^6}} \,
m_\chi\, \frac{\langle X^2\rangle}{M_{{\rm Pl}}^2}.
\end{equation}
Therefore, for $q=10^8$ and $m_\chi=10$, $\delta$ is of the order of
$3\times 10^{8} \langle X^2\rangle/M_{{\rm Pl}}^2$. Since
the final baryon asymmetry scales approximately as $\Gamma^{-1}$ and
is given by $B\simeq 5\times
10^{-4}\:\delta\:\epsilon\:(\Gamma/5\times 10^{-5})^{-1}$ \cite{klr},
where $\epsilon$ is an overall parameter accounting for
$CP$ violation (it will be typically a one-loop factor times some
CP-violating phases), we see that the observed baryon asymmetry
$B\simeq 4\times 10^{-11}$ may be explained by the phenomenon of GUT
baryogenesis after preheating if
\begin{equation}
\frac{\langle X^2\rangle}{M_{{\rm Pl}}^2}\simeq 5\times 10^{-13}
\left(\frac{10^{-2}}{\epsilon}\right)\left(\frac{\Gamma}{5\times 10^{-5}}
\right).
\end{equation}
From Fig.\ 3 we can read that this only may happen if $\Gamma_X$ is
smaller than about $10^{-3} m_X$.  This result may be considered very
comfortable since we can conclude that whenever the resonance
develops, i.e., when $\Gamma_X {\ \lower-1.2pt\vbox{\hbox
{\rlap{$<$}\lower5pt\vbox{\hbox{$\sim$}}}}\ }10^{-1} m=10^{-2} m_X$,
GUT baryogenesis after preheating is so efficient that the right
amount of baryon asymmetry is produced for almost the entire range of
values of the decay rate $\Gamma_X$. In other words, provided that
superheavy $X$-bosons are produced during the preheating stage, they
will be {\it ineffective} in producing the baryon asymmetry {\it only}
if their decay rate falls in the range $10^{-3} m_X {\
\lower-1.2pt\vbox{\hbox {\rlap{$<$}\lower5pt\vbox{\hbox{$\sim$}}}}\ }
\Gamma_X {\ \lower-1.2pt\vbox{\hbox
{\rlap{$<$}\lower5pt\vbox{\hbox{$\sim$}}}}\ } 10^{-2} m_X$.

%%%%%%%%%%%%%%%%%%%%%%%%%%%%%%%%%%
%%%%%%%%%%%%%%%%%%%%%%%%%%%%%%%%%%
%%%%%%%%%%%%%%%%%%%%%%%%%%%%%%%%%%
\begin{figure}
\centering
\leavevmode\epsfysize=3.6in \epsfbox{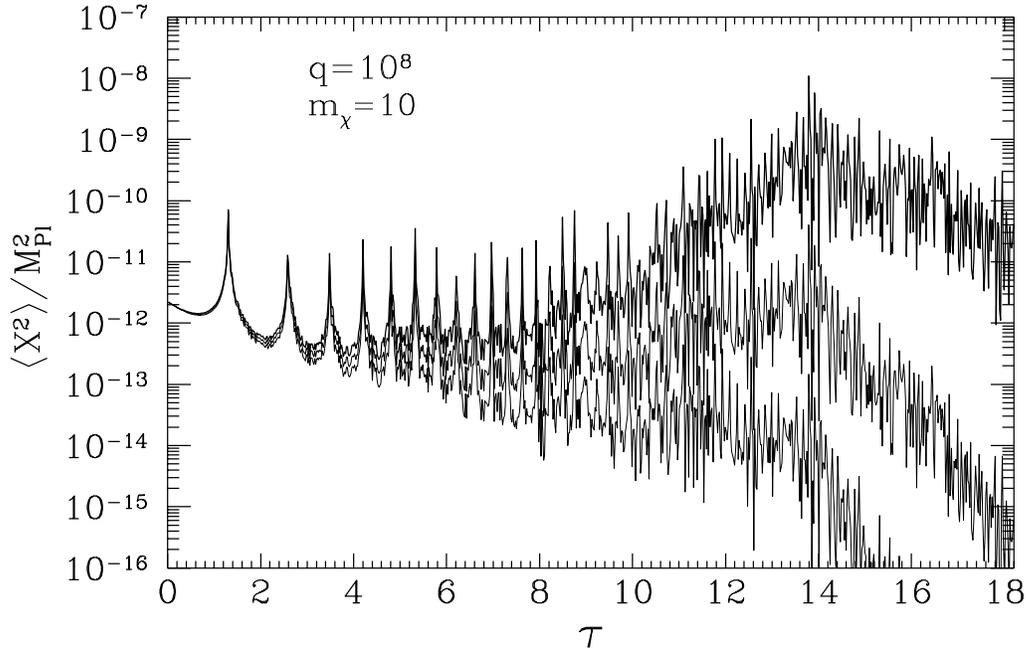}
\caption{The time dependence of the variance of $X$ in the Hartree 
approximation with
model parameters $q=10^8$,
 $m_\chi=10$ and for three values of $\Gamma$, from top to bottom:
 0.04, 0.08, 0.12.}
\label{fig:Fig3}
\end{figure}
%%%%%%%%%%%%%%%%%%%%%%%%%%%%%%%%%%
%%%%%%%%%%%%%%%%%%%%%%%%%%%%%%%%%%
%%%%%%%%%%%%%%%%%%%%%%%%%%%%%%%%%%

\section{Conclusions}
From our findings we may infer some phenomenological implications for
any model of GUT baryogenesis in preheating.  As we already mentioned,
$B$ violation through sphaleron transitions are expected to be fast at
high temperatures \cite{sh}, and would erase the baryon asymmetry
produced after preheating unless the supermassive $X$ bosons generate
some nonvanishing value of $B-L$.  A natural way to overcome this
problem is to adopt a GUT like $SO(10)$, where an asymmetry in $B-L$
may be generated by the lepton-number violating decays of the Higgs
field $\phi_{{\bf 126}}$ in the ${\bf 126}$-representation of
$SO(10)$ which transforms as a singlet under the $SU(5)$
decomposition.

This Higgs field is responsible for the Majorana masses of the
right-handed neutrinos through the symmetric couplings
$h_{\nu_R}\phi_{{\bf 126}}\cdot \nu_R^T C \nu_R$. Requiring that
$\Gamma_{\phi_{{\bf 126}}}$ is smaller than about $10^{-3}
m_{\phi_{{\bf 126}}}$ imposes the constraint $h_{\nu_R}{\
\lower-1.2pt\vbox{\hbox {\rlap{$<$}\lower5pt\vbox{\hbox{$\sim$}}}}\ }
10^{-1}$.  Through the see-saw mechanism, light neutrinos may
therefore get masses in the range $10^{-3}$ to $10^{-2}$eV for
$h_{\nu_R}\sim 10^{-3}$.

Notice that, as opposed to large-angle scattering processes,
forward-scatterings do not alter the distribution functions of the
particles traversing a gas of quanta, but only alter the dispersion
relation. This remains true even in the case of a nonequilibrium system
such as the one represented by the gas of $\phi_{{\bf 126}}$
particles created in preheating. At the end of reheating the right-handed 
neutrinos receive
a mass of the order of $h_{\nu_R} \langle \phi_{{\bf 126}}^2(1)
\rangle^{1/2}$ from non-thermal corrections with the $\phi_{{\bf
126}}(1)$-particles due to forward-scatterings. However, this mass
is not large enough to suppress kinematically the decay rate of the
$\phi_{{\bf 126}}$. This would require the energy of the
$\phi_{{\bf 126}}$'s to be smaller than $h_{\nu_R} \langle
\phi_{{\bf 126}}^2(1)\rangle^{1/2}$, which only happens if $\left(
\langle \phi_{{\bf 126}}^2(1)\rangle/M_{{\rm Pl}}^2\right)$ is larger
than about $10^{-8}\: g\: h_{\nu_R}^{-2}$.

In conclusion, we have performed a numerical analysis of the
production of superheavy $X$-bosons during the preheating stage
following the end of inflation and have shown that the observed baryon
asymmetry may be produced in the decay of these non-thermal GUT bosons
if the value of their decay rate is smaller than about
$10^{-3}m_X$. GUT baryogenesis after preheating solves many of the
serious drawbacks of GUT baryogenesis in the old theory of reheating
where the production of superheavy states after inflation was
kinematically impossible. Moreover, the out-of-equilibrium condition
is naturally attained in our scenario since the distribution function
of the $X$-quanta generated at the resonance is far from a thermal
distribution.  This situation is considerably different from the one
present in the GUT thermal scenario where superheavy particles usually
decouple from the thermal bath when still relativistic and then decay
producing the baryon asymmetry.

It is quite intriguing that out of all possible ways the parametric
resonance may develop, Nature might have chosen only those ways
without instantaneous thermalization and also with a successful
baryogenesis scenario.

\section*{Acknowledgements}

The work of EWK is supported in part by the Department of Energy and
by NASA grant NAG 5-2788. The work of IIT is supported in part by the U.S. Department of Energy
under Grant DE-FG02-91ER40681 (Task B) and by the National Science
Foundation under Grant PHY-9501458.  IIT would like to thank the Theoretical
Astrophysics Group at Fermilab and the CERN Theory Division where part
of this work was done for their kind hospitality.

\end{document}